# Research on Quality Measurement and Dynamic Transmission of Higher Education in China from the Perspective of Dual Circulation: An Empirical Analysis Based on Entropy Weight-TOPSIS and PVAR Models

Zhangming Yang [1], Xiaoting Yin [2], Dahai Zhang [3] *

**ABSTRACT**

Drawing on the new pattern of 'dual circulation', an evaluation system for the development level of China's higher education has been constructed, utilizing data from Chinese provinces over the period 2020 to 2022. Initially, the evaluation model is established, followed by the screening of initial indicators through correlation analysis. Subsequently, an entropy weight-TOPSIS model are employed for static quantification, and the PVAR model is introduced for the first time to describe the dynamic transmission from educational resources input to teaching effectiveness output. The empirical analysis reveals that, for effective development at this stage, higher education institutions should continuously optimize their educational structures, resource allocations, and talent cultivation strategies to align with the requirements of the new development paradigm, thereby contributing to national economic development through the provision of high-quality education.

**KEYWORDS**

“Dual circulation”, Entropy weight method, TOPSIS method, PVAR model, Dynamic transmission path

## 1. INTRODUCTION

Accelerating the construction of the 'dual circulation' development pattern is an important strategic plan formulated by China for the 14th Five-Year Plan and long-term development. In this context, China's higher education has taken the initiative to adapt to economic and social development, with the aim of gradually creating a new development pattern in which domestic and foreign markets can reinforce each other, with the domestic market serving as the mainstay [1]. This paper provides a comprehensive evaluation of China's current higher education development level, analyzing the development direction and reform results of China's higher education."

In fact, the "dual circulation" and higher education are integrated and mutually reinforcing. As for research on measuring the level of high-quality development of Chinese higher education under the "dual circulation" pattern, the existing literature mainly focuses on the following three aspects. The first aspect is theoretical research on the concept of "dual circulation". This includes theoretical analysis of the new development pattern. Qun-hui Huang, Xiao-juan Jiang, et al. believed that "dual circulation" is a proactive strategic choice for "cultivating new opportunities and opening up a new situation" in the context of the most profound and unprecedented changes the world has undergone in a century [2][3]. Yang Ge and Zi-xiang Yi put forward the view that "dual circulation" is an effective practice of Marx's theory of socialist reproduction [4]. The second aspect is the construction of indicators for high-quality development in higher education. This includes Hao-chang Yang et al.'s discussion and research on the construction of an evaluation index system for high-quality development of higher education in China [5]. Gunn A proposed the development of a teaching excellence framework as an indicator and method to measure the quality of teaching in higher education [6]. Horan W et al. established a practical framework with the sustainable development of higher education to assess selected international indicators [7], etc. Thirdly, there is theoretical research on higher education in the context of the "dual circulation" paradigm. This includes: From the aspects of structural system, scientific and technological innovation, opening up to the outside world, and governance capacity, Li Jie drew conclusions about optimizing the development pattern of higher education in the context of "dual circulation" [8]. Dun-rong Bie proposed the important impact of popular higher education on both the domestic and international levels from the perspective of "dual circulation" [9]. Shuai Jiang, Hai-long Li, and others conducted theoretical research on the changes and development of higher education under the new development pattern of "dual circulation" [10][11], etc. The research of these scholars has, to a certain extent, promoted the construction and development of the evaluation system for China's higher education development level under the "dual circulation" pattern.

This study constructs a quality evaluation framework for higher education under the "dual-circulation" development paradigm. At the theoretical level, grounded in the strategic logic of the dual-circulation development strategy and China's practical characteristics of higher education advancement, a hierarchical evaluation system was established through three dimensions: educational resources, student development support, and teaching effectiveness. This framework comprises 3 first-level indicators and 9 second-level indicators, encompassing critical elements such as material base, financial support, and Teaching staff strength. Methodologically, an innovative comprehensive evaluation model was developed by integrating the entropy weight method with the TOPSIS technique, through which a higher education quality index calculation formula was formulated to conduct provincial-level empirical measurements and comprehensive ranking analysis. The findings demonstrate that under the dual-circulation framework, the higher education system achieves synergistic development of quality enhancement and structural adaptation through optimized allocation of domestic and international resources, strengthened integration of scientific research and education, and reconstructed talent cultivation ecosystems.

## 2. CONSTRUCTION OF AN EVALUATION INDEX SYSTEM FOR HIGH-QUALITY HIGHER EDUCATION

### 2.1 Selection of evaluation indicators and construction of evaluation system

On the basis of literature research and China's 14th Five-Year Plan, this paper outlines how universities serve local needs under China's "dual circulation" model. Table 1 presents three primary indicators and nine secondary indicators:

**2.1.1 *Educational resources.***

Under the primary indicators of educational resources, this study delineates five secondary indicators: Material Base, Funding Support, Teaching Staff Strength, Disciplinary strength and Enrolment Scale. The degree of development in higher education is positively correlated with the robustness of the Material Base and the adequacy of Funding Support. The intensity of funding serves as the standard to measure the level of support for higher education in conducting scientific and technological activities and material expenditures in the region. The Disciplinary strength indicator is pivotal in assessing the internal diversity and the high-quality, advanced development of higher education, emphasizing the breadth and academic prowess of the teaching disciplines offered. Furthermore, the expansion of the Enrollment Scale is recognized as a significant factor in increasing revenue and enhancing profitability, as well as in cultivating excellent talents. This is also an important evaluation criterion established in this research. Based on the above, these indicators have reflected the foundation and comprehensive strength of colleges and universities, and are also the primary drivers for promoting internal circulation within the higher education ecosystem.

**2.1.2 *Student development and support.***

The outer circulation aspect of the 'dual circulation' paradigm is reflected in the integration of higher education with excellent international advanced practices. This paper summarizes the specific content of 'student development and support' into two secondary indicators: Internationalization of Talent Cultivation and International scientific and Technological exchange. Our approach, closely adhering to the principles of the 'dual circulation' policy, is designed to foster student development and promote internationalization. International talents serve as strategic resources for countries to enhance their comprehensive national strength and international competitiveness, thereby fully highlighting the interplay between the internal and external dimensions.

**2.1.3 *Teaching effectiveness.***

This paper encapsulates the concept of 'Teaching Effectiveness ' into two critical dimensions: Knowledge Production and Scientific Research Output. It is observed that universities consistently contribute a significant share to scientific research output and knowledge generation, underscoring their pivotal role in foundational research and the development of groundbreaking innovations. This reflects the vital contributions made by universities to national scientific and technological progress, and is also a key factor in promoting dual circulation within the economy.

Tables 1. Evaluation System for the Development Level of China's Higher Education Quality within the 'Dual Circulation' Paradigm

| Level 1 indicators A | Secondary indicators B | Tertiary indicators C | causality |
|---|---|---|---|
| A1 | B1 Material base | C1 School-owned Building Area | forward |
| | | C2 Intrinsic assets | forward |

| Level 1 indicators A | Secondary indicators B | Tertiary indicators C | causality |
|---|---|---|---|
| Educational resources | | C3 Proportion of Multimedia Classrooms in Higher Education Networks Percentage (%) | forward |
| | | C4 Teaching and Research Equipment Assets | Forward |
| | B2 Funding Support | C5 Special Allocation from Government Departments | forward |
| | | C6 Research and Development Project Funding | forward |
| | B3 Teaching Staff Strength | C7 Number of Full-time Teachers | forward |
| | | C8 Number of Professors and Associate Professors | forward |
| | B4 Disciplinary strength | C9 Number of National First-class Undergraduate Majors | forward |
| | B5 Enrollment Scale | C10 Number of Graduate Student Enrollments | forward |
| | | C11 Number of Undergraduate Enrollments | forward |
| A2 Student development and Support | B6 Internationalization of Talent Cultivation | C12 Number of Dispatched Personnel for Collaborative Research | forward |
| | | C13 Number of recipients of collaborative research | forward |
| | B7 International scientific and Technological exchange | C14 Number of Papers Presented at International Academic Conferences | forward |
| | | C15 Number of Hosted International Academic Conferences | forward |
| A3 Teaching Effectiveness | B8 Knowledge production | C16 Number of Awards for scientific and technological achievements | forward |
| | | C17 Number of Published Scientific and Technological Books | forward |
| | | C18 Number of Publication of academic papers | forward |
| | B9 Scientific Research Output | C19 Number of Invention Patent Applications | forward |
| | | C20 Number of Granted Invention Patents | forward |
| | | C21 Number of Patent Sale Contracts | forward |
| | | C22 Actual Income of the Current Year | forward |

*Actual Income of the Current Year is the amount of money received from the sale of the patent.

### 2.2 Data sources

Given the inherent lag in the compilation of annual statistical data, this paper selects the annual data from 2020 to 2022. The primary sources of these data are the “China Compendium of Science and Technology Statistics of Higher Education Institutions” (2020-2022)and “China Education Statistical Yearbook” (2020-2022）.

## 3. MODEL BUILDING

This study introduces an enhanced version of the TOPSIS method, which incorporates the entropy weighting approach. By adopting the entropy weight model to calculate the weight of each indicator, we can better eliminate the subjectivity in determining these weights and make the evaluation results more objective and fair. The entropy weighting method is an objective assignment method; the smaller the degree of variability among the indicators, the less information they reflect, and thus the lower the corresponding weight [12]. The TOPSIS method, on the other hand, can fully utilize the information in the original data to accurately reflect the differences between the evaluation programs. In contrast, the hierarchical analysis model tends to be too subjective in evaluation, and it is recommended that the number of decision-making factors should not exceed 10. By using TOPSIS, we can objectively determine the weights among multiple factors. Firstly, the entropy weight TOPSIS method is applied to evaluate the development level of China's higher education. Specifically, the entropy weight method is used to calculate the weights of the original data, and then the TOPSIS method is applied to analyze the newly weighted data to arrive at the final evaluation system. The specific analysis steps are outlined as follows.

### 3.1 Data preprocessing

The raw matrix is normalized to transform the indicator values into a consistent scale. The matrix is standardized in order to eliminate the impact of differing indicator scales, encompassing n provinces and m evaluation indicators, thereby obtaining the normalized matrix：

$$X = \begin{bmatrix} X_{11} & X_{12} & \cdots & X_{1m} \\ X_{21} & X_{22} & \cdots & X_{2m} \\ \vdots & \vdots & \ddots & \vdots \\ X_{n1} & X_{n2} & \cdots & X_{nm} \end{bmatrix} \tag{1}$$

Normalizing the forward matrix *X* results in the normalized matrix *Z*. The normalization formula applies to each element in *Z* as follows:

$$z_{ij} = x_{ij} / \sqrt{\sum_{i=1}^{n} x_{ij}^2} \tag{2}$$

The standardized matrix *Z* is formed as $Z = (Z_{ij})_{n*m}$, where $Z_{ij}$ represents the evaluation value of the *i*-th province under the *j*-th indicator.

$$Z = \begin{bmatrix} Z_{11} & Z_{12} & \cdots & Z_{1j} \\ Z_{21} & Z_{22} & \cdots & Z_{2j} \\ \vdots & \vdots & \ddots & \vdots \\ Z_{i1} & Z_{i2} & \cdots & Z_{ij} \end{bmatrix} \tag{3}$$

### 3.2 Entropy weight method of empowerment

In the first step, determine the value of the indicator for $P_{ij}$ :

$$P_{ij} = \frac{Z_{ij}}{\sum_{i=1}^{n} Z_{ij}}, (j = 1, 2, \cdots, m) \tag{4}$$

In the second step, the entropy value of the *j*-th indicator is calculated as $E_j$. If $P_{ij} = 0$ , $P_{ij} ln P_{ij} = 0$.

$$E_j = -\frac{1}{ln\, n} \sum_{i=1}^{n} P_{ij} ln\, P_{ij}, (j = 1, 2, \cdots, m) \tag{5}$$

($-\frac{1}{\ln n}$ in ln n, n is the number of objects, i.e. the number of rows, see matrix X)

In the third step, the coefficient of variation of item *j* is calculated as $G_J$:

$$G_J = 1 - E_j \tag{6}$$

In the fourth step, determine the weight of the *j*-th indicator $W_j$ :

$$W_j = \frac{G_J}{\sum_{j=1}^{m} G_J} \tag{7}$$

In the fifth step, calculate the entropy weight of each index based on $W_j$

### 3.3 Weighted-topsis model calculations

In the first step of the TOPSIS method, data normalization is performed to create a normalization matrix. This normalization matrix is then used as the basis for determining the weights of the attributes using the entropy weight method.

$$W_{ij} = \begin{bmatrix} W_{11} & W_{12} & \cdots & W_{1j} \\ W_{21} & W_{22} & \cdots & W_{2j} \\ \vdots & \vdots & \ddots & \vdots \\ W_{i1} & W_{i2} & \cdots & W_{ij} \end{bmatrix} \tag{8}$$

In the second step, a weighted normalized decision matrix V is constructed, where the elements $V_{ij}$ are calculated as $V_{ij} = W_j Z_{ij}$. Here, $W_j$.represents the weight of the *j*-th criterion, and $Z_{ij}$represents the normalized value of the i-th alternative on the j-th criterion.

$$V = \begin{bmatrix} V_{11} & V_{12} & \cdots & V_{1m} \\ V_{21} & V_{22} & \cdots & V_{2m} \\ \vdots & \vdots & \ddots & \vdots \\ V_{n1} & V_{n2} & \cdots & V_{nm} \end{bmatrix} = \begin{bmatrix} D_1Z_{11} & D_2Z_{12} & \cdots & D_mZ_{1m} \\ D_1Z_{21} & D_2Z_{22} & \cdots & D_mZ_{11} \\ \vdots & \vdots & \ddots & \vdots \\ D_1Z_{n1} & D_2Z_{n2} & \cdots & D_mZ_{nm} \end{bmatrix} \tag{9}$$

In the third step, determine the positive and negative ideal solutions.

Positive Ideal Solution:

$$V^{+} = (V_1^{+}, V_2^{+}, \cdots, V_m^{+}) = \{maxV_{ij} | j = 1, 2, \cdots, m\} \tag{10}$$

Negative Ideal Solution:

$$V^{-} = (V_1^{-}, V_1^{-}, \cdots, V_m^{-}) = \{minV_{ij} | j = 1, 2, \cdots, m\} \tag{11}$$

In the fourth step, the Euclidean distance is used to calculate the distance of the evaluation object from the positive and negative ideal solutions.

$$S_i^{+} = \sqrt{\sum_{j=1}^{m} \left(V_j^{+} - V_{ij}\right)^2, i = 1 \ldots n} \tag{12}$$

**3.4 Draw the results**

We calculate the relative proximity of each evaluation object (university) to the positive ideal solution, using the distances to both the positive and negative ideal solutions. This allows us to obtain a comprehensive evaluation of the educational quality at each university.

$$C_i = \frac{S_i^{-}}{S_i^{+} + S_i^{-}} \tag{13}$$

Where $0 \leq Ci \leq 1$, the larger Ci is, the higher it is.

## 4 EMPIRICAL ANALYSIS

### 4.1 Calculation of indicator weights using entropy weight method

Using the aforementioned formula, the weights of the indicators for each of the three levels of high-quality development of higher education in China's 'dual circulation' paradigm were calculated for the 31 provinces, spanning from 2020 to 2022.

Tables 2. Ratio of weights of indicators at three levels

| Tertiary Indicators | Weight Ratio | Rankings | Tertiary Indicators | Weight Ratio | Rankings | Tertiary Indicators | Weight Ratio | Rankings |
|---|---|---|---|---|---|---|---|---|
| C1 | 4.32% | 21 | C9 | 4.60% | 8 | C17 | 4.37% | 16 |
| C2 | 4.36% | 18 | C10 | 4.47% | 13 | C18 | 4.50% | 12 |
| C3 | 4.23% | 22 | C11 | 4.33% | 20 | C19 | 4.52% | 10 |
| C4 | 4.43% | 15 | C12 | 4.52% | 10 | C20 | 4.54% | 9 |
| C5 | 4.77% | 5 | C13 | 4.78% | 4 | C21 | 4.88% | 2 |
| C6 | 4.74% | 6 | C14 | 4.71% | 7 | C22 | 4.89% | 1 |
| C7 | 4.35% | 19 | C15 | 4.84% | 3 | | | |
| C8 | 4.37% | 16 | C16 | 4.47% | 13 | | | |

*The table presents the calculation result for the year 2021

As depicted in Table 2, international talent cultivation and scientific-technological exchanges carry the greatest weight among the tertiary indicators, reflecting China's accelerating educational internationalization under the dual-circulation framework. The university adopts a student-centered approach, encouraging participation in overseas study and exchange programs while securing high-quality global resources. It is further expanding its global engagement to advance regional internationalization [13] and build a high-level system for cultivating international talent.

Among research outputs, annual actual income receives the highest weight, confirming that integrating science with education is essential to translate universities' abundant research resources into talent-cultivation advantages and to drive the supply-side reform of human capital.

**4.2 Evaluation results based on the topsis method**

The relative proximity of each province to the optimal university has been calculated using a specific formula, yielding a comprehensive score and ranking for the period from 2020 to 2022. The evaluation results for this period are presented in Table 3.

Tables 3. Results of the Evaluation of High-Quality Development of Higher Education in Provinces

| Provinces | 2020 | | 2021 | | 2022 | | Three years | |
|---|---|---|---|---|---|---|---|---|
| | score | ranking | score | ranking | score | ranking | Average score | ranking |
| Jiangsu | 0.7459 | 1 | 0.7365 | 1 | 0.7206 | 1 | 0.7343 | 1 |
| Beijing | 0.6449 | 2 | 0.6128 | 2 | 0.6697 | 2 | 0.6425 | 2 |
| Guangdong | 0.5017 | 3 | 0.5626 | 3 | 0.5557 | 3 | 0.5400 | 3 |
| Shanghai | 0.4893 | 4 | 0.486 | 4 | 0.5145 | 4 | 0.4966 | 4 |
| Shandong | 0.4291 | 5 | 0.4542 | 5 | 0.4614 | 6 | 0.4482 | 5 |
| Hubei | 0.4188 | 6 | 0.4343 | 7 | 0.4616 | 5 | 0.4382 | 6 |
| Shanxi | 0.4007 | 8 | 0.4453 | 6 | 0.4346 | 7 | 0.4269 | 7 |
| Zhejiang | 0.3772 | 9 | 0.4128 | 8 | 0.4192 | 8 | 0.4031 | 8 |
| Sichuan | 0.4071 | 7 | 0.3788 | 9 | 0.4029 | 9 | 0.3963 | 9 |
| Henan | 0.3221 | 10 | 0.333 | 11 | 0.3161 | 11 | 0.3237 | 10 |
| Hunan | 0.2997 | 12 | 0.3339 | 10 | 0.3284 | 10 | 0.3207 | 11 |
| Liaoning | 0.3004 | 11 | 0.321 | 12 | 0.284 | 12 | 0.3018 | 12 |
| Anhui | 0.2459 | 13 | 0.2784 | 13 | 0.2754 | 13 | 0.2666 | 13 |
| Heilongjiang | 0.2441 | 14 | 0.2335 | 14 | 0.2362 | 14 | 0.2379 | 14 |
| Hebei | 0.2068 | 15 | 0.2184 | 15 | 0.219 | 15 | 0.2147 | 15 |
| Chongqing | 0.2002 | 16 | 0.2155 | 16 | 0.209 | 16 | 0.2082 | 16 |
| Jiangxi | 0.1843 | 19 | 0.2048 | 17 | 0.2008 | 18 | 0.1966 | 17 |
| Fujian | 0.188 | 17 | 0.1851 | 19 | 0.2067 | 17 | 0.1933 | 18 |
| Tianjin | 0.1848 | 18 | 0.1926 | 18 | 0.1884 | 19 | 0.1886 | 19 |
| Jilin | 0.1637 | 20 | 0.1806 | 20 | 0.165 | 21 | 0.1698 | 20 |
| Guangxi | 0.142 | 21 | 0.1799 | 21 | 0.1707 | 20 | 0.1642 | 21 |
| Shanxi | 0.1352 | 22 | 0.1394 | 22 | 0.1607 | 22 | 0.1451 | 22 |
| Yunnan | 0.1306 | 23 | 0.1318 | 23 | 0.1307 | 23 | 0.1310 | 23 |
| Guizhou | 0.103 | 24 | 0.107 | 25 | 0.1105 | 24 | 0.1068 | 24 |
| Gansu | 0.0955 | 25 | 0.1042 | 26 | 0.0908 | 26 | 0.0968 | 25 |
| Xinjiang | 0.0679 | 27 | 0.1114 | 24 | 0.109 | 25 | 0.0961 | 26 |

| Provinces | 2020 | | 2021 | | 2022 | | Three years | |
|---|---|---|---|---|---|---|---|---|
| | score | ranking | score | ranking | score | ranking | Average score | ranking |
| Inner Mongolia | 0.0887 | 26 | 0.0904 | 27 | 0.0904 | 27 | 0.0898 | 27 |
| Tibet | 0.0401 | 29 | 0.0572 | 28 | 0.0498 | 28 | 0.0490 | 28 |
| Qinghai | 0.0419 | 28 | 0.0415 | 29 | 0.033 | 31 | 0.0388 | 29 |
| Hainan | 0.0326 | 31 | 0.0364 | 30 | 0.0429 | 29 | 0.0373 | 30 |
| Ningxia | 0.0326 | 30 | 0.0338 | 31 | 0.034 | 30 | 0.0335 | 31 |

As illustrated in Table 3, the development level of education in China exhibits an uneven distribution, with Jiangsu, Beijing, and Guangdong ranking prominently. Jiangsu, in particular, has achieved an average score of 0.7343. Conversely, Qinghai, the Tibet, and the Ningxia lag significantly behind other provinces. The data obtained readily indicate that for local institutions of higher learning to achieve efficient development, they ought to continually optimize their educational structures, resource allocations, and talent cultivation in alignment with the demands of the new development paradigm. This optimization should aim to enhance the quality of education and thereby contribute to serving national economic and social development.

To enhance the interpretability of the results, this study plotted a line graph (not a "line graph table") based on the data from Table 3, as illustrated in Figure 1. The empirical analysis reveals that the evaluation of inter-provincial higher education levels in the past is roughly consistent with the factor dimensions in the measurement system of this paper. Municipalities with a relative concentration of higher education resources are more advantageous under the factor-oriented approach, resulting in Shanghai being ranked relatively higher than Guangdong and Jiangsu in previous studies. However, the indicator system established in this study is more closely aligned with the intrinsic demands of the "dual circulation" paradigm. It not only considers the "supply and demand" relationship required by domestic circulation, such as "factors of production" (education resources) and "labor achievements" (teaching effectiveness) in the higher education field within the indicator system, but also takes into account the mutual promotion between domestic and international circulations. Additionally, it introduces concepts such as local internationalization and the cultivation of high-level international talents in the development and support of students.

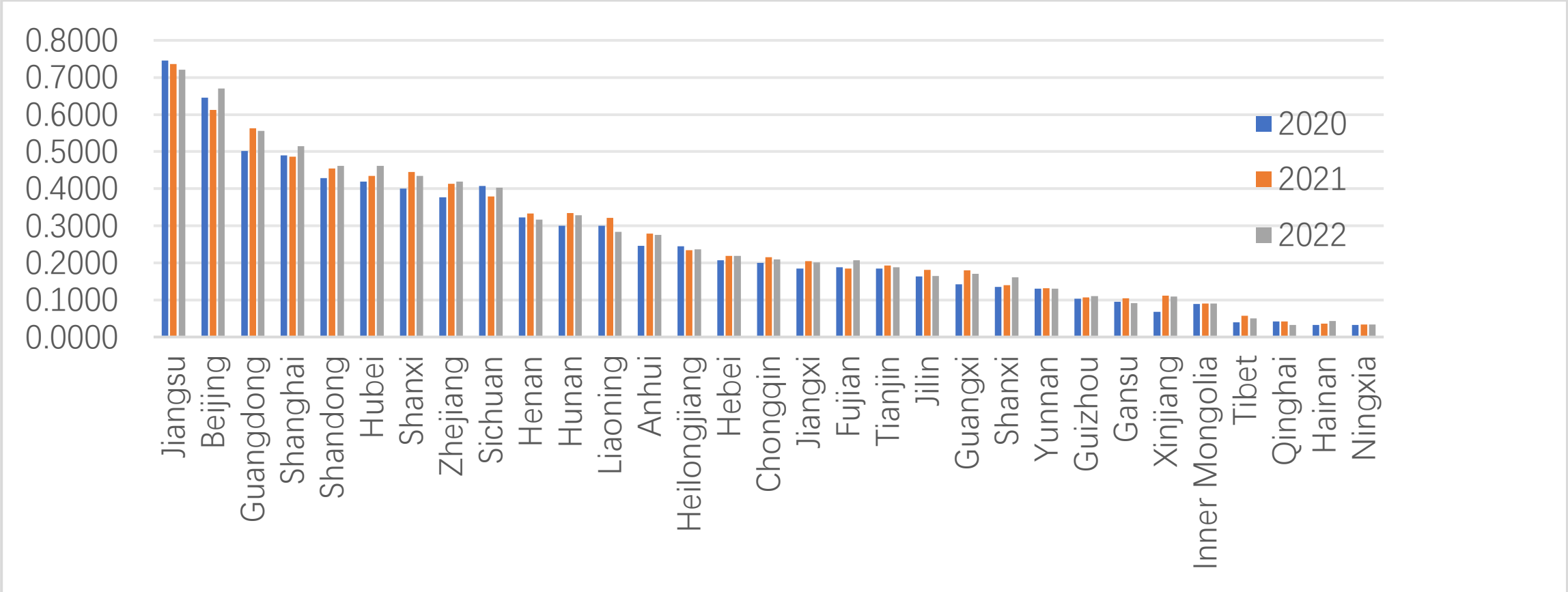

Figure 1. Line Chart of Provincial Higher Education Quality Development Scores

To address the potential bias of ad-hoc score-range classification, we employed the Jenks Natural Breaks algorithm (Jenks, 1967) to endogenously determine the cut-off points for the four development levels. The algorithm minimizes within-class variance and maximizes between-class variance among the 31 provincial composite scores. The resultant thresholds replace the original equal-interval divisions and are reported in Table 4.

Table 4. Category of Regional Comprehensive Development Levels of Higher Education in China

| Provincial Regions | Comprehensive Ranking | Score Range | Category |
|---|---|---|---|
| Jiangsu, Beijing, Guangdong, Shanghai | 1-4 | 0.649~0.734 | high-quality development |
| Shandong, Hubei, Shaanxi, Zhejiang, Sichuan, Henan, Hunan, Liaoning | 5-12 | 0.367~0.649 | relatively mature |
| Anhui, Heilongjiang, Hebei, Chongqing, Jiangxi, Fujian, Tianjin, Jilin, Guangxi, Shanxi | 13-22 | 0.159~0.367 | ascending phase |
| Yunnan, Guizhou, Gansu, Xinjiang, Inner Mongolia, Tibet, Hainan, Ningxia, Qinghai | 23-31 | 0.033~0.159 | underdeveloped |

The provincial regions categorized as high-quality development encompass Jiangsu, Beijing, Guangdong, and Shanghai. These regions exhibit a high level of comprehensive development in higher education, forming a highly harmonious interactive relationship with the economic and social development of their respective areas. Specifically, these four regions consist of two municipalities directly under the central government of China (Beijing and Shanghai) and two economically advanced provinces (Jiangsu and Guangdong). Leveraging their unique political status or economic prowess, these areas have actively devoted themselves to advancing higher education. Through a series of initiatives such as deepening international cooperation and exchanges, intensifying efforts to cultivate talent with an international perspective, and introducing advanced international educational resources, these regions have effectively enhanced the international influence of their higher education institutions, gradually establishing themselves as exemplary models of high-quality development in China's higher education landscape.

There is are eight relatively mature provincial-level administrative regions, namely Shandong, Hubei, Shaanxi, Zhejiang, Sichuan, Henan, Hunan, and Liaoning. These provinces have demonstrated notable achievements in the comprehensive development of higher education and played a pivotal role in facilitating the construction of China's internal economic circulation system. Nevertheless, there exists a degree of misalignment in the development of higher education within these regions, leading to an uneven contribution to the advancement of both China's internal and external economic circulations. Despite this, these areas continue to constitute the most emblematic sources of educational resource outputs in mainland China and serve as the mainstay supporting China's higher education as it advances towards a phase of high-quality development.

There are ten provincial-level administrative regions in the ascending phase, namely Anhui, Heilongjiang, Hebei, Chongqing, Jiangxi, Fujian, Tianjin, Jilin, Guangxi, and Shanxi. Although these provinces have demonstrated a certain level of emphasis on the high-quality development of higher education, they still exhibit notable gaps and deficiencies when compared to the requirements of the new development paradigm for optimizing educational structures, allocating resources reasonably, innovating talent cultivation models, and enhancing social contribution.

There are nine provincial-level administrative regions categorized as underdeveloped, namely Yunnan, Guizhou, Gansu, Xinjiang, Inner Mongolia, Tibet, Hainan, Ningxia, and Qinghai. The development of higher education institutions in these provinces has mainly followed traditional layouts. With the continuous growth of the regional economy, the gap between these institutions and those in other provinces in China has gradually emerged and tended to widen. This phenomenon underscores the urgent need for some higher education institutions to optimize their "dual-circulation" development planning model, namely the coordination mechanism between internal and external circulations, in order to adapt to the pressing demands under the new development paradigm of "dual-circulation" and accordingly promote in-depth teaching reforms and practices.

#### 4.3 Dynamic Causal Analysis

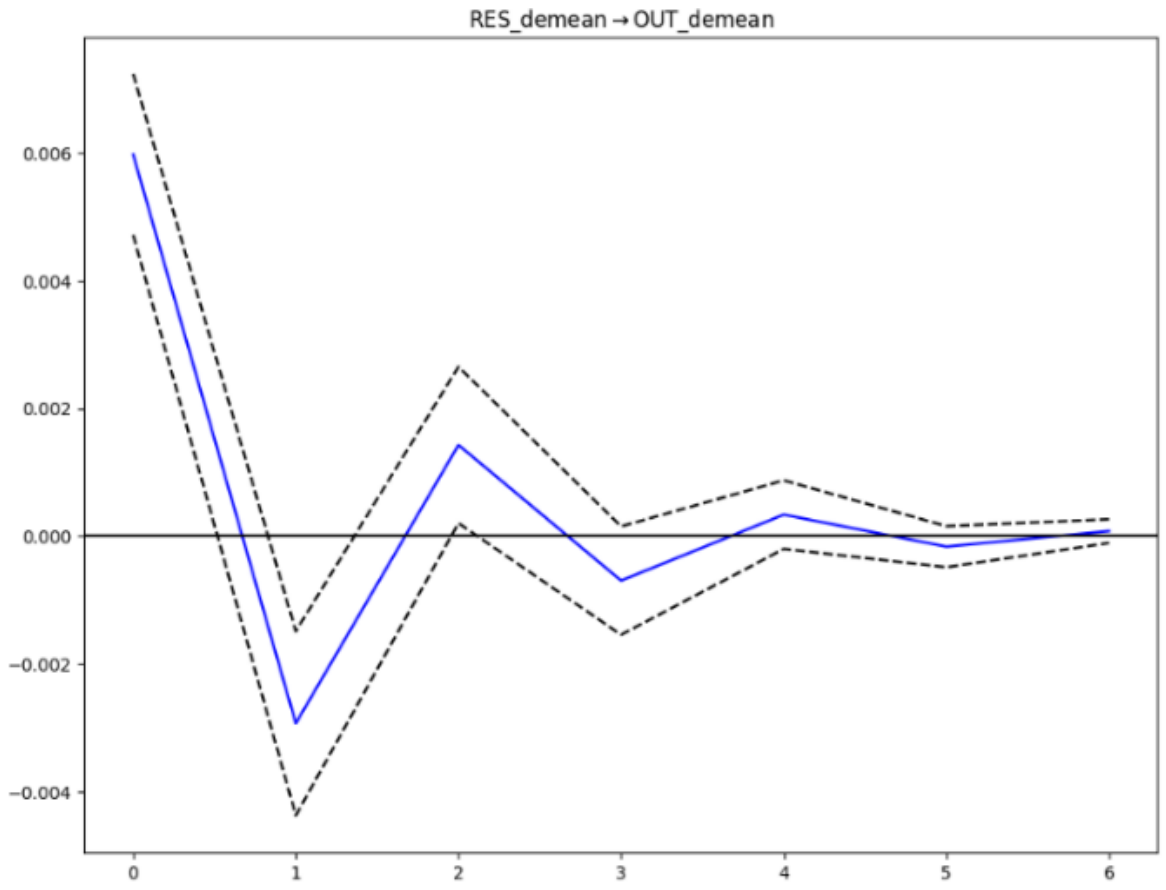

Figure 2. Impulse response function graph

Figure 2 presents the orthogonalized impulse-response functions from the PVAR model, illustrating how teaching-resource inputs (RES) dynamically influence teaching effectiveness outputs (OUT) under the “dual-circulation” framework. The shock is transmitted immediately: in period 0, additional RES lifts OUT, and the effect peaks at 0.006 standard deviations in period 1. Although the response gradually declines thereafter, it remains positive throughout, confirming that resource injections deliver a sustained boost to scientific output. Most of the impact is concentrated in the first two periods, a pattern consistent with the lengthy input–output cycle typical of higher education. Consequently, the findings carry clear policy implications: under the dual-circulation strategy, a continuous rise in educational investment can translate into short-term gains in research productivity, furnishing timely evidence that universities are effectively supporting the nation’s drive for scientific self-reliance.

## 5 CONCLUSIONS

### 5.1 Summary of conclusions

This paper establishes an evaluation system for assessing the high-quality development level of China's higher education within the framework of the "dual circulation" paradigm. The research findings emphasize three key points: Firstly, China places significant emphasis on charting a new course for internationalization within the context of the "dual circulation" framework. Secondly, the integration of science and education has emerged as a pivotal aspect of the ongoing human resources supply-side reform. Thirdly, higher education institutions should optimize their educational structure, resource allocation, and talent cultivation and training to align with the requirements of the "dual circulation" development pattern.

### 5.2 Policy Recommendations

Based on the empirical findings and aligned with current governmental and societal imperatives, this study proposes the following policy recommendations to advance the high-quality development of China's higher education system:

(1) Implement differentiated regional support policies, prioritizing provinces with educational resource disparities by enhancing targeted central fiscal transfers and establishing collaborative mechanisms for digital education infrastructure development and sharing.

(2) Refine incentive mechanisms for science-education integration, leveraging tax incentives and industry-university-research collaboration platforms to accelerate the effective translation of academic research outcomes into human capital supply.

(3) Develop a dual-circulation compatibility monitoring framework, integrating institutional evaluation metrics such as internationalization performance and dynamic alignment of disciplinary structures into university assessment systems.

These recommendations aim to address structural imbalances, foster innovation-driven resource allocation, and ensure systemic responsiveness to the evolving demands of the dual-circulation development paradigm.

**5.3 Future Research Directions**

Future studies should focus on three pivotal dimensions to deepen the exploration: First, expanding the spatiotemporal scope of datasets to trace the long-term educational effects of the dual-circulation strategy. Second, incorporating emerging indicators such as digital governance capabilities and cross-border educational collaboration to enhance the dynamic adaptability of the evaluation system. Third, conducting transnational comparative research, particularly examining synergistic development pathways among higher education systems in Belt and Road Initiative countries, thereby laying theoretical foundations for constructing a global-local dual synergy model in educational development.

**ACKNOWLEDGMENTS**

This work was supported by grants from the Liaoning Provincial Education Science Planning Leading Group Office (Project No. JG22DB073), the Liaoning Provincial Federation of Social Sciences (Project No. 2024lslybkt-020), and the Chinese Association of Higher Education (Project No. 23BR0203).

**REFERENCES**

[1] Fang F, Zhong BL. Theoretical Logic and Realistic Thinking of High-Quality Development of Higher Education under the New Pattern of “Dual Circulation” [J]. China Higher Education Research, 2022, (01):21-27.
[2] Huang QH. A "Dual Circulation" Development Pattern: Profound Connotation，the Background and Suggestions[J]. Journal of Beijing Institute of Technology (Social Science Edition) ,2021,21(01):9-16.
[3] Jiang XJ, Meng LJ. Mainly Inner Circulation, Outer Circulation Empowerment and Higher Level Double Circulation：International Experience and Chinese Practice. Management World[J],2021,37(01):1-19.
[4] Ge Y, Yin ZX. A Theoretical Analysis of my country’s Construction of “Double Cycle” New Development Pattern. Economic Issues[J],2021, (04):1-6.
[5] Yang H, Ge HZ Invention. Discussion on the construction of evaluation index system for high quality development of higher education[J]. Education Guide,2020, (10):83-90.
[6] Andrew Gunn. Metrics and methodologies for measuring teaching quality in higher education: developing the Teaching Excellence Framework (TEF)[J]. Educational Review, 2018, 70(2): 129-148.
[7] Development pattern under the establishment of "dual circulation" [J]. Higher Education Management,2021,15(05):23-35.
[8] Bie DR. The Significance of Popularization of Higher Education from the Perspective of “Dual Circulation” Strategy in China [J]. China Higher Education Research,2021, (05):22-28+35.
[9] Jiang S, Liu MC. The Reform and Development of Higher Education Under the New Development Pattern of “Double Circulation” Logic, Challenges and Proposals [J]. Heilongjiang Higher Education Research,2022,40(08):21-25.
[10] Li HL. The challenges of the “double cycle” pattern and the countermeasures of higher education[J]. Higher Education Management,2021,15(03):1-11.
[11] CHENG QY. Structural entropy weighting method for determining the weights of evaluation indexes[J]. Systems Engineering Theory and Practice,2010,30(07):1225-1228.
[12] Zhang W, Liu BC. Internationalization at home：New Trend of Higher Education in China[J]. University Education Science, 2017, (03):10-17+120.
[13] Jiang BQ. Internationalization at Home: An Ideal Option for Newly-Built Local Universities[J]. Journal of Hunan Institute of Humanities and Science,2016,33(02):99-102.